\documentclass[a4paper,12pt]{article}

\usepackage{latexsym}

\newcommand{\ud}{\mathrm{d}}

\begin{document}

\title{Metric gravity theories and cosmology. I. Physical 
interpretation and viability}

\author{Leszek M. SOKO\L{}OWSKI \\
Astronomical Observatory and Centre for Astrophysics,\\
Jagellonian University,\\ 
Orla 171,  Krak\'ow 30-244, Poland \\uflsokol@th.if.uj.edu.pl} 

\date{}
\maketitle

\centerline{Short title: Nonlinear gravity in cosmology}

\begin{abstract}
We critically review some concepts underlying current applications of 
gravity theories with Lagrangians $L=f(g_{\mu\nu}, R_{\alpha\beta\mu\nu})$ 
to cosmology to account for the accelerated expansion of the universe. 
We argue that one cannot reconstruct the function $f$ from astronomical 
observations either in cosmology or in the solar system. The 
Robertson--Walker spacetime is so simple and "flexible" that any cosmic 
evolution may be 
fitted by infinite number of various Lagrangians. We show on the 
example of Newton's gravity that one cannot recover the correct equation 
of motion from its approximate solution. Any gravity theory different 
from general relativity generates a new cosmological theory and all the 
successes of the standard cosmological model are lost even if a single 
solution of the theory well fits the observations. 
Prior to application 
of a given gravity theory to cosmology or elsewhere it is necessary to 
establish its physical contents and viability. This study may be performed 
by a universal method of Legendre transforming the initial Lagrangian in 
a Helmholtz Lagrangian. In this formalism Lagrange equations of motion 
are of second order and are the Einstein field equations with additional 
massive spin--zero and spin--two fields. All the gravity theories differ 
only by a form of interaction terms of the two fields and the metric. 
Initial conditions for the two fields in the gravitational triplet depend 
on which frame (i.e., the set of dynamical variables) is physical 
(i.e.~matter is minimally coupled in it). This fact and the multiplicity 
of possible frames obstruct confrontation of solutions to equations of 
motion with the observational data. A fundamental and easily applicable 
criterion of viability of any gravity 
theory is the existence of a stable ground state solution being either 
Minkowski, de Sitter or anti--de Sitter space. Stability of the ground 
state is independent of which frame is physical. 
\end{abstract}


\section{Introduction: problems with gravity theories} 
Metric theories of gravity, where the Lagrangian is any smooth 
scalar function of the curvature tensor, $L=f(g_{\mu\nu}, R_{\alpha\beta
\mu\nu})$, named metric nonlinear gravity (NLG) theories, have first 
attracted attention as possible candidates for foundations of quantum 
gravity due to their renormalizability properties \cite{Stel1} and then as 
a possible source of inflationary evolution of the very early universe 
\cite{Starob}. A recent revival of interest in these theories has come 
from cosmology. In fact, the theoretical state of affairs in cosmology is 
astonishing. The universe consists 
of baryons (4 percent), unknown stable massive particles forming nonbaryonic 
dark matter which do not fit the standard model of particle physics 
(26 percent) and dark energy (70 percent) about which we have 
only negative knowledge: these are not particles. About 96 percent of the 
material content of the universe is a great mysterious puzzle. 

On the other hand the laboratory experiments and astronomical observations 
confirming general relativity are still not very numerous and belong to a 
rather narrow class of tests. It is therefore attractive to conjecture that 
both the gravitational stability of galaxy clusters and the acceleration of 
the universe are not due to some unknown forms of matter, but can be accounted 
for by some modification of gravity theory. Modifications may go in all 
possible directions; here we deal with the most popular concept, the metric 
NLG theories. Among these the restricted NLG theories, wherein the Lagrangian 
is a function of the curvature scalar alone, $L=f(R)$, have been most 
frequently investigated. This approach to the dark matter and dark energy 
problem is sometimes referred to as a "curvature quintessence scenario". A  
typical motivation underlying this approach is following. Consider a 
Lagrangian of the form $L= R +R^2+\frac{1}{R}$ and the Robertson--Walker 
(R--W) 
spacetime. In the very early universe, when the curvature was large, the 
$R^2$ term was dominating generating some kind of inflation \cite{Starob}. 
At present $R$ is small and the $\frac{1}{R}$ term dominates giving 
hopefully rise to the accelerated evolution. And for the most time in the 
history the curvature scalar had intermediate values so that the linear 
term was leading preserving all the successes of the standard Friedmann 
cosmology \cite{Carr1, Carr2, NO1, NO2, CST, Cem, BBH}. This argument, 
attractive as it sounds, is misleading for three reasons. 

Firstly, if one makes a correction to the Einstein--Hilbert Lagrangian 
in the form $L=R+\varepsilon(R)$ where $\varepsilon(R)$ is any nonlinear 
function, it is not true that the resulting corrections to solutions of 
Einstein's theory are small when $\varepsilon(R)$ is very small and become 
significant only when $\varepsilon(R)$ is sufficiently large. The point is 
that any nonlinear correction to $L=R$ drastically alters the dynamical 
structure of the theory: the field equations become of fourth--order 
instead of second order and the higher curvature terms, even seemingly 
small, are always very important. To show the effect we consider a very 
simple model, a one--dimensional harmonic oscillator perturbed by 
introducing a small term with the third derivative, 
 \begin{displaymath}
\varepsilon \stackrel{...}{x}+\ddot{x}+\omega^2 x=0,
\end{displaymath} 
with $|\varepsilon|\ll1$. One seeks for solutions of the form $x=e^{\lambda t}$, 
then $\lambda = \textrm{const}$ is a solution of a cubic equation 
$\varepsilon \lambda^3 +\lambda^2 +\omega^2 =0$. One solves it by perturbing 
the two unperturbed solutions, i.e.~one sets $\lambda_{\pm} =\pm i\omega 
+\varepsilon\alpha_{\pm}$. Up to terms linear in $\varepsilon$ the solutions of 
the cubic equation are $\alpha_{\pm}=\frac{\omega^2}{2}$. Thus one has 
two almost periodic solutions 
 \begin{displaymath}
x_{\pm}(t) =a_{\pm}\exp\left[\left(\pm i\omega+\frac{\omega^2}{2}
\varepsilon\right)t\right]
\end{displaymath} 
being slowly damped or amplified periodic solutions for the unperturbed 
oscillator. These are approximations to exact solutions which are analytic 
in $\varepsilon$ at $\varepsilon=0$. However there is also a 
non-analytic at $\varepsilon =0$ solution of the 
differential equation corresponding to a third root of the cubic equation. 
Assuming that $\lambda_3$ is of order $\varepsilon^{-1}$ and keeping only the 
leading terms (of order $\varepsilon^{-2}$) one gets $\lambda_3 =-\frac{1}
{\varepsilon}$ and the third solution is exponentially growing ($\varepsilon <0$) 
or fading, 
\begin{displaymath}
x_3 =a_3\exp(-\frac{t}{\varepsilon}).
\end{displaymath} 
This solution is qualitatively different from the other two and exists for 
arbitrarily small $\varepsilon$. One expects that the space of solutions is 
significantly extended by qualitatively new solutions due to any correction 
$\varepsilon(R)$ to $L=R$. 

Secondly, the Lagrangian is not a physical observable whose value or 
variability gives some insight into the state of a given physical system. 
It is a kind of a generating function giving rise to equations of motion 
for the system and observables such as energy--momentum tensors 
(canonical and variational). Any gravitational Lagrangian is, 
by definition, made up of scalars, in general these 
are all invariants of the Riemann tensor (and possibly their derivatives), 
while the resulting field equations 
are tensor ones. Any assumption about the value of a scalar appearing in $L$, 
say $R$, actually tells very little about corresponding solutions. For 
instance, setting $R=0$ in general relativity one gets not only all vacuum 
solutions but also those for matter with a traceless energy--momentum tensor. 
Actually one can say something nontrivial about the sought for solution 
merely by inspection of the Lagrangian $L=f(R)$ only in the case of the 
simplest non--maximally symmetric spacetime, the Robertson--Walker one. In fact, 
the Riemann tensor for this metric is determined by the cosmic scale factor 
$a(t)$ and for any gravitational Lagrangian (also that explicitly depending 
on the Weyl curvature) the field equations reduce to one quasilinear third 
order ODE for $a(t)$, hereafter named \emph{the quasi--Friedmannian 
equation\/}. Then assuming 
that $R$ is large in some epoch of cosmic evolution and small in another one, 
one may neglect small terms in this equation and find approximate (or even 
exact) solutions in these epochs. In so doing one must take care since there 
is no easily available information about values of $R$ in various eras of the 
cosmic evolution. In fact, in general relativity $R=-8\pi G T$ where $T$ is 
the trace of the matter energy--momentum tensor and for cosmic perfect fluid 
$T=3p-\rho$. In the early universe the ultrarelativistic plasma has the 
equation of state $p=\frac{1}{3}\rho$ and then $R=0$. Hence in the 
transition from the radiation to the matter era $R$ increases, contrary to 
what has usually been assumed in the motivation for $L= R +R^2+\frac{1}{R}$ 
and only later on it slowly fades. What is more important, the relationship 
between $R$ and $T$ (as well as between $R_{\mu\nu}$ and $T_{\mu\nu}$) 
is algebraic only in general relativity and for a 
nonlinear $L=f(R)$ it becomes a differential equation for $R$. If the 
field equations (in Jordan frame) are $E_{\mu\nu}(g)=8\pi G T_{\mu\nu}$, 
then their trace reads 
\begin{displaymath}
Rf'(R) - 2f(R) +3 f'''(R)g^{\mu\nu}R_{,\mu}R_{,\nu} +3 f''(R)
\Box R =8\pi G T
\end{displaymath} 
and one cannot a priori estimate which terms in the equation are dominant 
and which ones may be neglected. To this end one must assume not only an 
approximate value of $R$ in a given epoch but also that $R$ is slowly and 
monotonically varying in order to estimate the derivatives; in this way one 
eliminates possible rapidly oscillating solutions. All this 
makes sense if one is convinced that 
he deals with the correct Lagrangian, e.g.~the Lagrangian has been derived 
from first principles (string theory, quantum gravity etc.) or otherwise 
motivated. Going in the opposite direction, i.e.~attempting, as is recently 
done, to reconstruct the underlying Lagrangian from observed qualitative 
features of the cosmic scale factor means that one has to construct the whole 
relativistic cosmology anew.\\

In fact, from the general Hawking--Penrose singularity theorem, valid in 
general relativity, it follows that our universe contains a singularity 
since the cosmic fluid satisfies the strong energy condition. Then in the 
case of the Robertson--Walker spacetime the Friedmann equation implies that 
the singularity was in the past, the initial Big Bang, and the cosmic scale 
factor monotonically grows from zero at the curvature singularity. We stress 
that these are generic qualitative properties of any solution to the 
Friedmann equation, independent of a specific equation of state for the 
cosmic fluid matter. On the other hand for a generic $L=f(R)$ the 
singularity theorem  cannot hold. Whether or not the theorem holds must 
be proved case by case\footnote{For theories in which the theorem holds 
it takes a form different from that in general relativity since 
$R_{\mu\nu}$ is not a function of $T_{\mu\nu}$.}. For 
an arbitrary function $f(R)$, even for many of those Lagrangians which 
admit a solution qualitatively fitting the astronomical data (an 
acceleration phase at present preceded by a deceleration phase), there was 
no initial singularity. There was no Big Bang and the following "early 
universe" when it was small, dense and hot. And the quasi--Friedmannian 
equation for such a Lagrangian does not ensure that the cosmic scale factor 
grows monotonically from an initial small value. What is actually done by 
those authors who say "today the $R^{-1}$ term is leading in $L$ while in the 
early universe the $R^2$ term was dominant" is merely choosing a class of 
(approximate) solutions for which $a$ and $R$ significantly and
monotonically vary in the 
course of a cosmic evolution; the solutions in this class are, by 
construction, ever 
increasing. This is a kind of "fine tuning" since one may expect that 
there are classes of 
qualitatively different solutions which are no less typical. This 
conjecture is at least partially supported by rigorous investigations in the 
Einstein frame where the third gravitational degree of freedom for $L=f(R)$ 
gravity is revealed as a nonlinear scalar field minimally coupled to Einstein 
gravity: for certain simple (power--law) scalar field potentials all 
solutions (in the spatially flat Robertson--Walker spacetime) are oscillatory 
at late times while for potentials which are bounded from above there exist 
solutions which are global to the past with the Hubble parameter $H = 
\dot{a}/a$ converging to a constant nonzero value as $t\to -\infty$ 
\cite{Ren}. Taking into 
account the complexity of the quasi--Friedmannian equation for any nonlinear 
$f(R)$, investigation of the qualitative properties of \emph{all} solutions 
is not easy and can only be done (case by case) by performing a 
phase--space analysis. Is is fortunate that the equation can be reduced 
to a first order equation for $\dot{H}$ viewed as a function of $H$ in the 
case of the Einstein--de Sitter spacetime ($k=0$) and a second order one 
for the open and closed universe\footnote{There is a subtle mathematical 
assumption necessary to decrease the order of the equation, namely that $H(t)$ 
is a monotonic function. Then the analysis loses all solutions $a(t)$ 
which are not monotonic.} (Starobinsky in ref. \cite{Starob}). A 
preliminary analysis of evolution of the Einstein--de Sitter metric in 
the case od a couple of special Lagrangians was performed by Carroll et 
al. \cite{Carr2}. For the Lagrangian most frequently studied, introduced 
in \cite{Carr1}, they found an attractor solution $a\to t^2$ for late 
times and an exact solution $a \propto t^{1/2}$ starting from a 
curvature singularity at $t=0$. However it is difficult to see from their 
phase portraits, which cover only a piece of one quadrant of the phase 
space, whether all solutions emerge from the singularity or there 
are nonsingular solutions (besides the exponentially growing and decreasing 
ones with constant $H$) and whether oscillatory (non--monotonic) solutions 
are excluded. It should be emphasized that if for a given accepted 
Lagrangian the two questions are 
answered "no", then all the successes of the standard Friedmann cosmology 
are lost.\\ 

The last statement leads us to the third reason, taking a form of a 
problem: to what extent can one reconstruct the Lagrangian from a given 
solution? Clearly any given function may be viewed as a solution of 
many diverse differential equations. Requirement that the function is 
a solution of a Lagrange equation imposes stringent restrictions on 
possible equations and allows one to look for a unique answer. 
A simple example shows that there are cases when it can be 
effectively done under some conditions. The Newton's gravitational force 
may be expressed as the gradient of a potential and one seeks for a 
corresponding Lagrangian. The potential generates a differential scalar 
$S=\partial_i\phi \, \partial_i\phi$ ($x_i$ are the Cartesian coordinates) 
and the kinetic part of any Lagrangian should be some function $f(S)$. 
Furthermore there may be a potential energy $V(\phi)$ and a candidate 
Lagrangian is $L= f(S)-V(\phi)$. Inserting the Newton's potential 
$\phi= -\frac{\alpha}{r}$ into the Lagrange equation 
\begin{displaymath}
4f''(S)\phi_{,ik}\phi_{,i}\phi_{,k} + 2f'(S)\triangle\phi +V'(\phi) =0
\end{displaymath} 
($\phi_{,i}\equiv \partial_i\phi$) one gets 
\begin{displaymath}
-\frac{8\alpha^3}{r^7} f''(S) +V'(\phi) =0.
\end{displaymath} 
The problem is indeterminate since there are two unknown functions and 
one equation. A unique solution corresponding to the Laplace equation 
arises either upon setting $V=0$ or requiring linearity of the 
resulting equation. \\
Suppose now one is attempting to determine the Lagrangian of an  
underlying restricted NLG theory from observations in the solar system. 
It is reasonable to assume that the metric is static spherically 
symmetric. As we discuss further in this work one of essential 
requirements that any viable $L=f(R)$ theory should 
satisfy is that it 
admits solutions either with $R_{\mu\nu}=0$ or $R_{\mu\nu}= \Lambda g_{\mu
\nu}$ with $\Lambda$ constant positive or negative (Einstein spaces); 
clearly there are 
infinitely many Lagrangians having this feature. Thus each $L=f(R)$ 
theory under consideration has either Schwarzschild or 
Schwarzschild--(anti)de Sitter spacetime as a solution and if it is this 
solution that is realized in the nature then no set of observations and 
experiments can reconstruct the function $f(R)$. In these theories the 
Birkhoff theorem does not hold and other static spherically symmetric 
solutions do exist (very few of them are known \cite{MV}). Any solution 
different from $R_{\mu\nu}= \Lambda g_{\mu\nu}$ is generated by nonzero 
values of a scalar field being a nongeometric component of a 
gravitational doublet of fields, see section 3. There are two possibilities 
depending on whether the doublet is in the Jordan frame (more precisely, 
in Helmholtz--Jordan frame) or in the Einstein frame. If the measurable 
quantities form the Jordan frame, then the spin--0 gravity has ordinary 
matter as a source. In the solar system the source (the sun) enforces 
spherical symmetry of the scalar. The field is massive and unless its 
mass is extremely small it is a short range one. The presence of the 
scalar gives rise to a non--Schwarzschildean solution which is very close 
to ordinary Schwarzschild one (possibly with $\Lambda\neq 0$). 
Nevertheless a recently found approximate solution for a specific form 
of $f(R)$ shows that even small corrections are 
detectable\footnote{This solution belongs to the class of solutions which 
are generated by the higher-than-second order of the field equations in 
JF, i.e.~it is analogous to the third solution for the perturbed 
harmonic oscillator discussed above. The solution does not reduce 
to a Schwarzschildean one in the limit $f(R)\to R$ and this is why it 
considerably deviates from the former even if the difference 
$f(R)-R$ is negligibly small.} and in this 
case they are ruled out by measurements \cite{ESK}. In the 
case the Einstein frame is observable the scalar is independent of any 
local matter distribution and in particular in the solar system it is 
not determined by position and mass of the sun. It is rather of 
cosmological origin, i.e.~the initial 
data are fixed and common for the entire universe.
Then the scalar field is 
(almost) homogeneous and to avoid a conflict with the cosmological 
observations it must be very weak giving rise to unmeasurable effects in 
the solar system. This means that all observations performed there  
will confirm the Schwarzschild metric. Even the 
cosmological constant is undetectable locally since the most sensitive 
to it effect, the perihelion shift of Mercury, requires $|\Lambda| \geq
 10^{-41}\,m^{-2}$ \cite{KKL}, while the cosmologically admissible value 
is $|\Lambda| \approx 10^{-52}\,m^{-2}$. Thus observations may effectively 
distinguish between the two frames. If Jordan frame is physical then small 
corrections to Schwarzschild metric generated by the scalar are ruled out 
by the observations in vicinity of the sun in the case of most 
Lagrangians investigated up to now \cite{Olm, CST, RI, BBH, APT, Ch1, 
CSE,CB1}, though there are arguments that the linear approximation 
applied to calculate these corrections may not be valid within the 
solar system \cite{NVA}.\\
 
 The Robertson--Walker spacetime has a higher (six--dimensional) symmetry 
 group and its metric contains an arbitrary function which may be freely 
 adjusted so that each metric NLG theory, i.e.~any 
 $L=f(g_{\mu\nu}, R_{\alpha\beta\mu\nu})$, admits it as a solution of  
 the corresponding quasi--Friedmannian equation and in 
 this sense this spacetime is "flexible" and universal while Minkowski, 
 de Sitter and anti--de Sitter spaces are not. Hence one expects that 
 infinitely many of the theories predict a deceleration phase 
 in the past and an accelerated evolution in the present epoch. It has  
 been conjectured that any cosmological evolution may be realized by some 
 specific $L=f(R)$ \cite{NO2}. And it is impressive that the 
 astronomical data may be fitted by such diverse functions as rational 
 and exponential ones \cite{DBD}, a combination of two confluent 
 hypergeometric functions \cite{NO2}, a combination of two 
 hypergeometric functions \cite{CDD} and 
 even an implicit form of $f(R)$ was found which is consistent with 
 the three years collection of WMAP data \cite{NO2}. Also a "toy model" 
 based on a minimal curvature conjecture ($R$ is always above some 
 minimal value $R_0>0$) fits exceptionally well the $\Lambda$CDM 
 cosmological model \cite{Ea}. These results were 
 found by studying the quasi-Friedmannian equation, i.e.~the dynamics of 
 $L=f(R)$ theory; if one applies the dynamically equivalent Einstein frame 
 it is easy to show that for any scale factor $a(t)$ there exists at least 
 one function $f(R)$ having this factor as a solution of the field 
 equation\footnote{Yet in a recent paper it is found that for a large 
 class of the Lagrangians the evolution of linear density fluctuations 
 does not fit the structure formation on both large and small scales 
 simultaneously \cite{BBPST}.} \cite{Woo}. Therefore there is 
 no surprise that also by inclusion of corrections in a form of inverse 
 squares of the Ricci and Riemann tensors the Supernovae data can be 
 fitted without the need of any dark energy \cite{MSW, Carr2}. On the 
 other hand in the metric--affine formalism of gravity theories it was 
 found that the current cosmological data (the supernovae Ia "gold set", 
 the CMBR shift parameter and the linear growth factor) impose stringent 
 restrictions on the test Lagrangian $L= R +\alpha R^{\beta}$ since 
 $\beta\approx 0,05$ \cite{AEMM}; this result, however, does not 
 uniquely confirm the $\Lambda$CDM model and is rather due to the specific 
 choice of the Lagrangian.\\
 
 The main problem of cosmology which arises in this field of 
 investigations is therefore whether it is possible at all to effectively 
 uniquely recover the underlying Lagrangian $L=f(R)$ from an exact 
 analytic solution $a(t)$ to the quasi--Friedmannian equation. 
 (Reconstruction of a Lagrangian depending on more than one curvature 
 invariant is impossible.) 
 The problem was investigated in more detailed way in two papers. 
 First Capozziello et al. in an ingenious work \cite{CCT} attempted to 
 determine $f(R)$ assuming that one knows from observations a precise 
 analytic dependence of the Hubble parameter $H$ on the redshift 
 parameter $z$, $H(z)$. They expressed the curvature scalar $R$ as a 
 function of $H(z)$ and then derived an equation for $f[R(z)]$ from 
 the quasi--Friedmannian equation\footnote{The evolution equation for 
 $a(t)$ may be transformed into an equation for $a(z)$ and then for 
 $f[R(z)]$ provided $z(t)$ is a monotonic function. As discussed 
 above this means that one takes into account only this class of 
 solutions. It is unclear as to what extent this restriction affects 
 the final outcome.}. The equation for $f$ is a linear third order 
 ODE with coefficients being extremely complicated functions of 
 $H(z)$ and hence, in spite of its linearity, is analytically 
 intractable. The authors were able to find out the initial data 
 from observations for the equation so that there is a unique 
 solution, but it is inaccessible. The method is unique and  
 ineffective. It requires an exact analytic form of $H(z)$ while 
 astronomic observations provide only a finite set of values 
 $H(z)$ which are affected by large errors. To circumvent this 
 difficulty the authors used a couple of simple functions 
 approximating the real dependence of $H$ on $z$ which arise in 
 recent models for dark energy such as quintessence and the 
 Chaplygin gas. Given these analytic expressions, the numerical 
 integration of the equation for $f[R(z)]$ is inevitable, then the 
 solution is approximated by an empirical function. The final 
 analytic fit for $f(R)$ looks rather unconvincing (eq. (48) of 
 \cite{CCT}) and the authors caution the reader against drawing any 
 physical implications from it. Recently  Fay et al. \cite{FNP} 
 have attempted to search for the form of $f(R)$ corresponding to the 
 $\Lambda$CDM cosmology applying the general autonomous system for 
 the $k=0$ R--W spacetime. They found, among other functions, 
 $L=R- 2\Lambda$, but this Lagrangian corresponds to saddle points 
 of the autonomous system while the attractors of the system do not 
 reproduce the radiation and the matter eras of the universe and 
 give rise to a complicated irrational function $f(R)$. Their work 
 clearly shows that the spatially flat R--W spacetime is particularly 
 deceptive in this respect. In fact, the dynamics of this spacetime 
 is not so simple as is usually assumed. We almost never employ the 
 \emph{exact\/} solutions of the Friedmann equation for the real 
 matter content of the universe as they are too involved and we are 
 so accustomed to the approximate solutions (which very well fit the 
 observations) that we view them as exact ones. For example, the 
 period when the energy densities of the "radiation" (all the 
 ultrarelativistic particles including neutrinos) and of the "dust 
 matter" (nonrelativistic baryons and possibly the dark matter) were 
 comparable was longer than the whole radiation era and in this period 
 $a(t)$ was very complicated, nevertheless it is commonly assumed that 
 in the radiation era there was no dust and this era was suddenly 
 transformed into the matter era and after the transition the 
 radiation disappeared. Under this assumption one gets the simple 
 expressions $t^{1/2}$ and $t^{2/3}$ for $a(t)$ in these eras 
 respectively; this is the basic assumption in \cite{FNP}.\\
 
 It should be stressed that in general it is impossible to reconstruct 
 a differential equation from its approximate solution. A given 
 function is an approximate solution to a whole class of equations and 
 the correct equation is in no way distinguished within this class: 
 the Einstein equations correspond to the saddle points rather than to 
 attractors of the dynamical system of ref. \cite{FNP}. Metric NLG 
 theories are so conceptually involved that this (rather obvious) fact 
 goes overlooked. As an example we take a simpler theory. Recall that 
 it is possible to reconstruct the Newtonian theory (Laplace 
 equation) from the exact solution, the Newton's potential. Now one 
 tries to reconstruct it from an approximate solution. Suppose an 
 astronomer observes a planet of mass $m$ moving on a very 
 elongated elliptic orbit with the eccentricity $e=1-
 \varepsilon$, $\varepsilon\ll 1$ and 
 $\varepsilon$ positive. He assumes that the 
 gravitational force comes from a central potential\footnote{In 
 conformity with classical mechnics we denote it by $U$ instead of 
 $\phi$.} $U(r)$ and tries to determine it from the planet's motion. 
 To this end he employs the energy integral 
 \begin{displaymath}
 E=\frac{m\dot{r}^2}{2} +\frac{J^2}{2mr^2} +U(r)
\end{displaymath} 
where $J$ is the conserved angular momentum. The astronomer 
determines from observations the positions $r$ at different times and 
fits the data by an analytic function $r(t)$, inverts it and 
expresses the derivative $\dot{r}$ as a function $F(r)$. He makes 
his observations in two regimes: when the planet is close to 
perihelion and to aphelion. From the exact solution of the Kepler 
problem we may compute that near the perihelion the dependence 
$t(r)$ for this elongated orbit is approximately (by keeping only 
the dominant term in $\varepsilon$)
\begin{displaymath}
 t\approx \frac{1}{3}\sqrt{\frac{2m}{\alpha}} r^{3/2}
\end{displaymath} 
for $U=-\frac{\alpha}{r}$ and this function should be used by the 
astronomer as an analytic fit to his data set. When the planet 
moves near the aphelion the approximate solution $t(r)$ which 
also fits the other observational data is 
\begin{displaymath}
 t\approx \sqrt{\frac{ma^3}{\alpha}} \bigg(\pi -4\sqrt{1-\frac{r}
 {2a}}\bigg).
\end{displaymath} 
Both the functions are easily invertible and each is used to 
compute $\dot{r}=F(r)$. The quantities $a$ (the semimajor axis), $e$, 
$J$ and the parameter $p$ are determined from the observations 
(assuming ellipticity of the orbit), then $\alpha =\frac{J^2}
{mp}$ and upon inserting $F(r)$ into the energy integral the 
astronomer finally finds that for $r$ close to $r_{\textrm{min}}$ 
\begin{displaymath}
 U(r)\approx -\frac{J^2}{mpr} - \frac{J^2}{2mr^2} + E
\end{displaymath}
while close to the aphelion
\begin{displaymath}
 U(r)\approx -\frac{J^2}{4mpa^2}(2a-r) -\frac{J^2}{2mr^2} + E.
\end{displaymath}
None of these expressions is an approximation to the Newton's 
potential. An ultimate conclusion from this example and the two 
works mentioned above is that reconstructing the underlying 
Lagrangian from known Robertson--Walker solutions is impossible for 
fundamental reasons. 
 Relativistic metric theories of gravity are too intricate. It is 
 in order here to remind 
 the Einstein's view that a new physical theory is never formulated 
 by induction from a pile of empirical data. \\
 
 All that above does not imply that alternative theories of gravity 
 should not be applied in cosmology and in particular should not be 
 used to account for the dark energy. Today the situation in 
 gravitational physics is exceptional as compared to other branches 
 of physics: the well established and confirmed theory, general 
 relativity, seems to be just a point in the "space" of all existing 
 and conceivable theories of gravitational interactions and its nearest 
 neighbourhood is densely populated by its alternatives, the metric NLG 
 theories. According to Cantor's theorem the cardinal number of the set 
 of metric NLG theories is greater than the cardinality of the 
 continuum. The very existence of these theories entitles one to apply 
 them to describe effects which are gravitational or may be interpreted 
 as such. On the other hand the wealth of these theories makes necessary, 
 before making any applications of one chosen from this huge set, to 
 investigate two problems: i) to determine all possible interrelations 
 between them and their relationships to general relativity, ii) whether 
 a given gravity theory satisfies all the well grounded general rules 
 of classical field theory and has acceptable properties. These 
 include: \\
 ---determination of particle contents (spectrum), \\
 ---existence of a stable maximally symmetric ground state,\\
 ---form of interactions with ordinary matter, i.e.~which quantities are 
 measurable. \\
 A natural and obvious postulate is to require that 
 a given gravity theory should have a
 Newtonian limit. It however creates conceptual problems. 
 It is not sufficient to formally get in the linear approximation the 
 Poisson equation for a scalar potential \cite{Di} 
 $\Delta U=4\pi G\rho+\Lambda$ or another equation 
 appearing in the Newtonian gravity (such as the stellar hydrostatic 
 equilibrium one \cite{BB}). Newtonian gravity is well defined as a 
 small static perturbation of Minkowski spacetime and the postulate 
 may be unambiguously formulated for those Lagrangians which admit the 
 flat spacetime as a solution. In this class of theories the postulate 
 may work as an effective discriminating condition. Yet the NLG theories 
 are most interesting and attractive for cosmologists if their 
 ground state solution is curved being de Sitter or anti--de 
 Sitter space while the flat spacetime is excluded. In this case the 
 very concept of Newtonian gravity in these spacetimes is controversial 
 and there are conflicting results on the weak--field limit for different 
 Lagrangians.
 As far as we know the problem of the Newtonian limit is not convincingly 
 solved even in general relativity in the presence of the cosmological 
 constant (a precise definition of the limit is lacking). Following 
 cosmologists we deal mainly with the theories 
 which do not admit Minkowski space as a solution and at the present 
 level of knowledge the Newtonian limit criterion cannot be applied 
 to them.\\
 
 Investigations of the second problem, i.e., to what extent the metric 
 NLG theories are viable from the viewpoint of classical field theory 
 are the main subject of this paper and the subsequent Paper II. 
 In section 2 we count the number of 
 degrees of freedom and briefly discuss the particle spectrum of a general 
 NLG theory. The latter subject requires replacing the fourth--order 
 Lagrange equations by dynamically equivalent second order ones for the 
 resulting gravitational triplet of fields; this aim is achieved by the 
 powerful method of Legendre transformations. This decomposition is 
 crucial for all investigations of the theory. The triplet may be described 
 in infinitely many different frames and the two most important ones and 
 the problem of coupling ordinary matter to the gravity fields are 
 presented in section 3. The controversial problem of which frame is 
 physical does not affect the criteria of viability of various gravity 
 theories. The most fundamental criterion is the existence of a stable 
 maximally symmetric ground state solution and it is investigated in 
 Paper II. 
 Conclusions and further considerations concerning the 
 possibility of recovering the underlying $L=f(R)$ Lagrangian from a 
 given (cosmological) solution are contained in section 4. 

\section{Particle spectrum}
We consider in this section the general metric NLG theory based on 
an arbitrary Lagrangian $L=f(g_{\mu\nu}, R_{\alpha\beta\mu\nu})$. The 
theory is metric in the sense that a nondegenerate tensor field 
$g_{\mu\nu}$ with Lorentzian signature is the only independent dynamical 
quantity. One may also investigate a metric--affine theory ("the 
Palatini method") with the same Lagrangian wherein one takes independent 
variations of $L$ with respect to the metric and a symmetric connection. 
As is well known, for any $L$ different from $R$ the two theories 
diverge. Some authors claim that the metric--affine approach is more 
natural since the Lagrange equations are of second order while in the 
purely metric theory they are of fourth order. However the metric NLG 
theory is not inherently a higher derivative one. The tensor $g_{\mu\nu}$ 
appearing in the Lagrangian actually is a kind of unifying field mixing 
various particles (fields) with different spins and masses. To find out 
a physical interpretation of the field it is necessary to decompose it 
in a multiplet of these fields. Then equations of motion for the 
separate fields are of second order and display a physical content of 
the theory better than the original fourth--order ones. This is why 
referring to NLG theories as "higher derivative ones" is misleading. 
In this 
respect the metric--affine approach is not advantageous over the purely 
metric formalism and we prefer the latter as conceptually simpler. \\

To avoid any confusion and for sake of completeness we begin with 
determining the degrees of freedom of the general metric NLG theory 
though it may be found in the literature. Counting the degrees of 
freedom (d.o.f.) in this case is far from trivial. The quadratic theory, 
$L=R+R^2+R_{\mu\nu}R^{\mu\nu}$, is known to have eight d.o.f. \cite{
Stel1, BD, FT, BCh}. For Lagrangians with arbitrary dependence on the 
Ricci tensor and applying a perturbative approach to Lagrangians 
depending on the Weyl tensor the d.o.f. were first 
counted by Hindawi et al. \cite{HOW1} by using a second order version 
of the theory. For most general Lagrangians it is possible to determine 
the maximal number of d.o.f. in the initial fourth--order formulation. 
It is well known that the pure gravitational field in general relativity, 
$R_{\mu\nu}=0$, has two d.o.f. \cite{Sa} and we shall count them in the 
same way for $L=f(g_{\mu\nu}, R_{\alpha\beta\mu\nu})$. The number of 
d.o.f. for a given system is defined as a half of the number of 
arbitrary functions needed to uniquely specify the initial data for the 
Cauchy problem for the equations of motion of the system. The equations 
of motion following from this Lagrangian form a system of ten tensor 
PDE of fourth order $E^{\alpha\beta}=0$ for the unifying field 
$g_{\mu\nu}$. Let a spacelike hypersurface S be chosen as an initial 
data surface for the equations. The theory is generally covariant 
(diffeomorphism invariant) and one can freely choose a coordinate system 
in the spacetime and the most convenient one is the comoving system 
(normal Gauss coordinates), $g_{00}=-1$, $g_{0i}=0$, such that S has 
an equation $t=0$. There are six unknown functions $g_{ik}$ and the 
initial Cauchy data consist of values of $g_{ik}$, $\partial g_{ik}/
\partial t$, $\partial^2 g_{ik}/\partial t^2$ and 
$\partial^3 g_{ik}/\partial t^3$ on S; these are 24 functions of 
three coordinates $x^i$. The data are subject to a number of constraints. 
First, the coordinates on S may be freely changed and this gives a 
freedom of choice of 3 functions of these variables. Secondly, the 
trace of the extrinsic curvature of S may be given any value, i.e.~one 
function is arbitrary. Finally, the general covariance of the theory 
implies, in the same way as in general relativity, that some of the 
field equations are constraints. In fact, the invariance of the 
action 
\begin{displaymath}
S=\int \ud^4x \sqrt{-g} f(g_{\mu\nu}, R_{\alpha\beta\mu\nu})
\end{displaymath} 
under an infinitesimal coordinate transformation gives rise to a strong 
Noether conservation law ("a generalized Bianchi identity") 
$\nabla_{\beta} E^{\beta}_{\alpha}\equiv 0$ (see e.g.~appendix A in 
arXiv version of \cite{MS1}). Explicitly it reads 
\begin{displaymath}
\partial_0 E^0_{\alpha} + \partial_i E^i_{\alpha} + \Gamma^{\beta}
_{\beta\nu} E^{\nu}_{\alpha} - \Gamma^{\nu}_{\beta\alpha} E^{\beta}
_{\nu} =0. 
\end{displaymath} 
The last three terms contain at most fourth time derivatives of 
$g_{\mu\nu}$ and the identity implies that $\partial_0 E^0_{\alpha}$ 
cannot involve fifth time derivatives. Thus $E^0_{\alpha}$ involve 
at most third time derivatives and $E^0_{\alpha} =0$ are not 
propagation equations but form four constraints on the initial 
data. Together the number of independent Cauchy data is diminished 
to 16 arbitrary functions and thus a general metric NLG theory has 
eight d.o.f. \\

For a restricted NLG theory, $L=f(R)$, 
the number of d.o.f. is less than 8 what means that the equations 
$E^{\alpha}_{\beta} =0$ generate additional constraints on the 
Cauchy data. It is not easy to determine all the constraints from these 
equations and one should instead apply a second order formulation of 
the theory. To this end it is adequate to view the  
$L=f(R)$ theories in a wider context of as large class of of NLG 
theories as possible. We therefore consider for the time being Lagrangians 
$L=f(g_{\mu\nu}, R_{\alpha\beta})$, i.e.~with no dependence on the 
Weyl tensor. \\
An adequate mathematical tool for this purpose is provided by a specific 
Legendre map \cite{MFF1, JK, MFF2}. For these NLG theories the method 
is as general and powerful as Legendre maps transforming the Lagrangian 
formalism into the Hamiltonian formalism in classical mechanics and 
classical field theory. Yet the method is not currently used in a 
systematic way and most papers on applications in cosmology have 
employed various \emph{ad hoc\/} tricks to transform from the Jordan 
frame to the Einstein frame. The tricks in most cases give results 
equivalent to the Legendre transformation, however do not allow to 
fully display the structure and features of the theory. Here we give 
a brief summary of investigations of particle spectrum contained in 
\cite{MS2} while the general formalism is described in \cite{MFF1, JK, 
MFF2}. \\

The Jordan frame consists of only one dynamical variable, the tensor 
field $g_{\mu\nu}$, $\mathrm{JF}=\{g_{\mu\nu}\}$, which plays both the 
role of a metric tensor on a spacetime $M$ and a kind of unifying 
gravitational field being a composition of some fields having definite 
spins and masses. Pure gravity is then described by a multiplet of the 
fields having together at most eight d.o.f., the metric is a geometric 
component of the multiplet. The unifying field may be decomposed into 
the component physical fields in two ways. The first method assumes 
that $g_{\mu\nu}$ is the spacetime metric and one separates from it, 
by means of a Legendre map, the additional degrees of freedom, i.e.~the 
other components of the multiplet. The Ricci tensor is decomposed into 
its irreducible parts, the trace $R$ and the traceless tensor 
$S_{\mu\nu} \equiv R_{\mu\nu}- \frac{1}{4}g_{\mu\nu}R$, then one defines 
a scalar and a tensor canonical momentum conjugate to the "velocity" 
$R_{\mu\nu}$ by Legendre transformations 
\begin{displaymath}
\chi +1 \equiv \frac{\partial L}{\partial R}, \qquad 
\pi^{\mu\nu}\equiv\frac{\partial L}{\partial S_{\mu\nu}}=\pi^{\nu\mu}.
\end{displaymath} 
Together with the metric the two fields form a gravitational triplet, 
named the Helmholtz--Jordan frame, $\mathrm{HJF}=\{g_{\mu\nu}, \chi, 
\pi^{\mu\nu}\}$. Equations of motion for the triplet follow from a 
Helmholtz Lagrangian $L_H$ which is dynamically equivalent to the initial 
$L=f(g_{\mu\nu}, R_{\alpha\beta})$. Here one meets a technical 
obstacle: to get an explicit form of $L_H$ one must express $R$ and 
$S_{\mu\nu}$ in terms of $\chi$ and $\pi^{\mu\nu}$, i.e.~to solve the 
defining equations to find $R=r(\chi, \pi^{\mu\nu})$ and 
$S_{\mu\nu}=s_{\mu\nu}(\chi, \pi^{\alpha\beta})$. For a general 
$L=f(g_{\mu\nu}, R_{\alpha\beta})$ this requires solving nonlinear 
matrix equations. (No doubt, the power of Hamiltonian formalism in 
physics stems from the fact that physically relevant Lagrangians are 
quadratic in "velocities".) This is why only Lagrangians that are 
quadratic in the Ricci tensor have been investigated in detail. 
However the formalism in principle works for any $L$. Furthermore, 
as we shall see, in the Einstein frame the specific dependence of 
$L$ on $R_{\mu\nu}$ only affects interaction terms while the general 
structure of the theory remains unaffected, therefore although 
a general 
formalism (in HJF) has been developed for any function $f$, in 
practice one applies only those $f$ for which the Legendre 
transformations may be effectively inverted. For any $f$ the 
Helmholtz Lagrangian reads 
\begin{equation}
L_H =R + \chi R + \pi^{\mu\nu}S_{\mu\nu} -H(\chi, \pi^{\mu\nu}),
\end{equation}
where $H$ is a Hamiltonian. This Lagrangian is linear in $R_{\mu\nu}$, 
what implies that Lagrange equations for $g_{\mu\nu}$ take form of  
Einstein field equations, $G_{\mu\nu}= 8\pi G T_{\mu\nu}(R_{\alpha\beta}, 
\chi, \pi^{\alpha\beta})$. The RHS of these equations is by definition 
an energy--momentum tensor for the two fields and it depends 
linearly on the Ricci tensor and on the first and second derivatives 
of $\chi$ and $\pi^{\mu\nu}$. There are no kinetic terms for $\chi$ 
and $\pi^{\mu\nu}$ in $L_H$ and propagation equations for the fields 
are derived in a rather intricate way from $T_{\mu\nu}$, these are 
hyperbolic second order ones. Both the fields are subject to one 
algebraic and four first order differential constraints and in 
consequence they carry together six degrees of freedom. In summary, 
the particle spectrum of the theory exhibited in HJF consists of: 
a massless spin--2 field (graviton\footnote{We use this traditional 
name for the metric field satisfying $R_{\mu\nu}=0$. It is worth noting, 
however, that the relationship between the hypothetical quantum of the 
gravitational field and the classical field (described by general 
relativity) is different from that between the photon and the classical 
Maxwell field, see \cite{SS}.}, spin two and 2 d.o.f.), a massive 
spin--2 field (5 d.o.f.) and a massive scalar field. This outcome 
(with the same values of the masses for the two fields) was first found 
in the linear approximation \cite{Stel1, Stel2} and then by various 
methods in the exact theory for a quadratic Lagrangian $R+ R^2 +
R_{\mu\nu}R^{\mu\nu}$ \cite{ABJT, HOW2}. HJF is not uniquely determined: 
since the Helmholtz Lagrangian in (1) is not in a canonical form, various 
redefinitions of $\chi$ and $\pi^{\mu\nu}$ are admissible. \\

Introducing the field $\pi^{\mu\nu}$ makes sense if its definition may 
be (at least in principle) inverted to yield 
$S_{\mu\nu}=s_{\mu\nu}(\chi, \pi^{\alpha\beta})$, otherwise 
$\pi^{\mu\nu}\equiv 0$. The field exists if the Lagrangian depends on 
$R_{\mu\nu}$ in a nontrivial way, i.e.~the Hessian 
\begin{displaymath}
\det\left(\frac{\partial^2 f}{\partial R_{\alpha\beta}\partial R_
{\mu\nu}}\right) \neq 0.
\end{displaymath}
This condition does not hold for $L=f(R)$. Then $\pi^{\mu\nu}$ vanishes 
and gravity is described by a doublet $\mathrm{HJF}=\{g_{\mu\nu}, \chi 
\}$ carrying 2+1 d.o.f. and the scalar cannot be massless. Thus for this 
class of Lagrangians determining the particle spectrum is very simple 
and straightforward. \\

The other approach to constructing a second order formalism is more 
sophisticated. Here one assumes that $g_{\mu\nu}$ is merely a unifying 
field for gravitation and plays a role of a spacetime metric in a 
purely formal way---in the sense that the Ricci tensor appearing in the 
Lagrangian is made up of it. One introduces a new metric as a canonical 
momentum conjugate to the full Ricci tensor via a Legendre map as 
\cite{MFF1, JK}
\begin{equation}
\tilde{g}^{\mu\nu} \equiv (-g)^{-1/2}\left|\det\left(\frac
{\partial f}{\partial R_{\alpha\beta}}\right)\right|^{-1/2}
\frac{\partial f}{\partial R_{\mu\nu}},
\end{equation}
here $g= \det (g_{\mu\nu})$. $\tilde{g}^{\mu\nu}$ may be viewed as a 
metric 
tensor providing that $\det(\partial f/\partial R_{\alpha\beta})\neq 
0$. Clearly for $f(R)=R$ one gets $\tilde{g}^{\mu\nu}=g^{\mu\nu}$ and 
for arbitrary $f(R)$ the new metric is conformally related to the old 
one, $\tilde{g}_{\mu\nu}= f'(R)\,g_{\mu\nu}$, where $\tilde{g}_{\mu\nu}$ 
is the matrix inverse to $\tilde{g}^{\mu\nu}$. In this case the 
Legendre transformation is degenerate since it cannot be inverted. 
In general the transformation is a map of a (Lorentzian) metric 
manifold $(M, g_{\mu\nu})$ into another one, $(M, \tilde{g}_{\mu\nu})$. 
If the transformation (2) is invertible, i.e.~the Hessian for $f$ does 
not vanish, it may be solved to give $R_{\mu\nu} = r_{\mu\nu}
(g_{\alpha\beta}, \tilde{g}^{\lambda\sigma})$. As in 
Helmholtz--Jordan frame one constructs a Helmholtz Lagrangian which 
now takes a generic form 
\begin{equation}
\tilde{L}_H = \tilde{R}(\tilde{g}) + K(\tilde{\nabla}g) -
\tilde{g}^{\mu\nu} r_{\mu\nu}(g, \tilde{g}) +\left|\frac
{\det(g_{\alpha\beta})}{\det(\tilde{g}_{\alpha\beta})}\right|^
{1/2} f(g_{\mu\nu}, r_{\alpha\beta}(g, \tilde{g})),
\end{equation}
where $\tilde{\nabla}$ is the covariant derivative with respect to 
$\tilde{g}_{\mu\nu}$. It is worth stressing that in Einstein frame, 
$\mathrm{EF}=\{\tilde{g}_{\mu\nu}, g_{\alpha\beta}\}$, precisely the 
Einstein--Hilbert Lagrangian for the spacetime metric 
$\tilde{g}_{\mu\nu}$ is recovered, giving rise to Einstein field 
equations $\tilde{G}_{\mu\nu}(\tilde{g})= 8\pi G \tilde{T}_{\mu\nu}
(\tilde{g}, g)$ with the tensor field $g_{\mu\nu}$ acting as a 
"matter" source for 
the metric. The tensor $\tilde{T}_{\mu\nu}$ is the variational 
energy--momentum tensor for $g_{\mu\nu}$ defined in the standard way 
and contains second derivatives $\tilde{\nabla}_{\mu}\tilde{\nabla}_
{\nu} g_{\alpha\beta}$ but no curvature\footnote{In general the 
energy--momentum tensor involves second derivatives of matter 
variables, the gauge fields, the minimally coupled scalar field and 
perfect fluids belong to few exceptions.}. Unlike $L_H$ in HJF, the 
Lagrangian has a canonical form for $g_{\alpha\beta}$, i.e.~is a 
sum of a kinetic term and a potential part. The kinetic term, 
$K(\tilde{\nabla}g)$, is a quadratic polynomial in first derivatives 
$\tilde{\nabla}_{\mu} g_{\alpha\beta}$, as is usual in classical field 
theory and what is really remarkable, it is \emph{universal\/}, 
i.e.~is independent of the form of the function $f$ \cite{MFF1, JK}. 
The only reminiscence of the original $L$ in JF is contained in the 
potential part of $\tilde{L}_H$: explicitly via $f(g_{\mu\nu}, 
r_{\alpha\beta})$ and implicitly via $r_{\alpha\beta}$, i.e.~in 
interaction terms\footnote{Clearly this is not little. In quantum 
mechanics every state vector satisfies the Schr\"odinger equation and 
the whole variety of quantum systems is encompassed in interaction 
terms in the Hamiltonian. Here something analogous occurs. We stress 
this point since in the fourth--order formulation of an NLG theory in 
JF an impression arises that the theory is more different from general 
relativity than it indeed is.}. Hence in EF 
one recovers just general relativity with a source which may be 
interpreted either as a nongeometric component of gravity or merely 
as a (quite exotic) matter field described by classical field theory. 
In this sense Einstein general relativity is a universal Hamiltonian 
image (under the Legendre map) of any $L=f(g_{\mu\nu}, R_{\alpha\beta})$ 
gravity theory. In other terms, general relativity is an isolated 
point in the space of all gravity theories: its closest neighbourhood,  
consisting of the metric NLG theories, can be mapped onto it and thus 
is not different from GR. 
It is also clear that in practice there is no need 
in studying Lagrangians more complicated than quadratic in $R$ and 
$R_{\mu\nu}$. \\

The second order Lagrange equations for $g_{\mu\nu}$ in EF are subject to 
four differential constraints (following from Bianchi identities 
$\nabla_{\nu} G^{\nu}_{\mu}(g) =0$) which allow one to eliminate 
four of ten components of the field. This shows that it carries six 
d.o.f. and is actually a mixture of two different physical fields. 
The next step is thus to decompose it into components with definite 
spins. Then one again gets a scalar and a spin--2 field. For practical 
purposes it is convenient to eliminate the scalar from the outset by 
an appropriate choice of the Lagrangian in JF \cite{MS2}. The field 
$\psi_{\mu\nu}$ arising in this way from $g_{\mu\nu}$ carries spin 
two (and five d.o.f.), has the same mass as that computed in HJF for 
$\pi^{\mu\nu}$ and is nonlinear (it is well known that any linear 
spin--2 field is inconsistent in general relativity \cite{AD}). It is 
straightforward to show in EF that $\psi_{\mu\nu}$ is necessarily a 
ghost field (a "poltergeist") \cite{Stel1, Stel2, ABJT, HOW2, Tom}, 
while it is rather difficult to establish this feature for 
$\pi^{\mu\nu}$ in HJF. And to avoid any misunderstanding we stress 
that the ghost--like behaviour of the spin--2 field is inevitable: 
it appears in any consistent theory of gravitationally interacting 
spin--2 fields \cite{Wa} and in particular is a feature of any 
$L=f(g_{\mu\nu}, R_{\alpha\beta})$ gravity. \\

All these gravity theories are similar and they differ only in the 
interaction terms in (3) and in masses of the spin--2 and scalar 
components of the gravitational triplet. All the $L=f(R)$ theories 
are reduced in Einstein frame to general relativity plus a massive 
minimally coupled field with a self--interaction potential determined 
by $f$ \cite{MS1}. At first sight it seems that Lagrangians 
generating tachyonic masses of the fields should be excluded as 
untenable. This is the case when the ground state solution is 
Minkowski space and other fields behave as small excitations in 
this spacetime. However, if the ground state is anti--de Sitter, 
also the scalar field with a tachyon mass is allowed if its 
modulus is not too large in comparison to the cosmological constant,  
see Paper II. \\

The case of most general Lagrangians, $L=f(g_{\mu\nu}, 
R_{\alpha\beta}, C_{\mu\nu\alpha\beta})$, explicitly depending on 
the Weyl tensor, is more subtle. The unifying field $g_{\mu\nu}$ 
carries 8 d.o.f. and one conjectures that it can be decomposed in 
the same triplet of the graviton, a spin--2 field and a scalar, 
however a proof for an arbitrary $L$ is missing. The conjecture was 
proved only in a perturbative analysis (and in four dimensions): 
one expands a generic Lagrangian about a ground state solution of 
the theory (Minkowski, de Sitter or anti--de Sitter) up to terms 
quadratic in Riemann tensor and arrives at \cite{HOW1, Ch2}
\begin{displaymath}
L= \mathrm{const} + R+ aR^2 +bR_{\mu\nu}R^{\mu\nu} +c\mathrm{GB},
\end{displaymath}
where GB, the Gauss--Bonnet term, is a topological invariant and 
in $d=4$ may be discarded as a divergence. Thus the most general 
NLG theory perturbatively reduces to that without the Weyl tensor 
and has the same particle spectrum. \\

We stress again that in both the frames any NLG theory is reduced to 
general relativity plus some exotic source. In EF the 
general--relativistic form of the theory is obvious, in HJF it is less 
conspicuous due to the specific form of the Helmholtz Lagrangian 
(1). From the physical viewpoint the spin--two and spin--0 fields 
may be viewed either as the components of the gravitational triplet 
or just as a kind of matter. It is not quite clear whether the 
difference between the two interpretations is empirical: whether 
there is a "gedankenexperiment" allowing to differentiate one from 
the other. As in Brans--Dicke theory it is necessary to assume 
that the two fields do not couple to any other matter in the sense 
that in a relevant Lagrangian there are no interaction terms of the 
two fields with the particles of the standard model. And as in 
Brans--Dicke theory "one way" interactions are admissible: ordinary 
matter may act as a source in equations of motion for $\chi$ and 
$\pi^{\mu\nu}$. The two nongeometric components of gravity only 
interact gravitationally in the sense that there is interaction 
between them and, first of all, they act as a source of the 
spacetime metric in Einstein field equations. It is often assumed 
(and clearly there is no proof) that the dark energy signals its 
existence solely by its influence on the cosmic evolution. Whether 
it should be regarded as a form of matter or as a component of 
gravity is presently a matter of convention. Mathematically the 
issue is irrelevant and Einstein field equations arising from (1) 
and (3) should be studied by applying all the methods developed 
to this aim in general relativity. In particular, the 
energy--momentum tensors for the spin--0 and spin--2 fields 
appearing in these equations ought to satisfy the conditions 
usually imposed on matter in general relativity. Clearly it may be 
claimed that the two components of gravity are specific in the 
sense that they need not satisfy the standard conditions, e.g.~the 
scalar field may not be subject to energy conditions. It might be 
so, however the price would then be high: the most significant 
results (or their appropriate analogues) found in general relativity 
would be inaccessible in a given NLG theory. In this work we assume 
that the spin--0 gravity is not exceptional in this respect. 

\section{Frames and initial conditions}
Once an NLG theory is expressed in HJF one may perform arbitrary 
Legendre transformations ("canonical ones") and changes of variables 
(field redefinitions), thus the theory may be formulated in infinitely 
many various frames. Clearly both HJF and EF are priviledged 
by their construction exhibiting the physical content of the theory. All 
the frames are mathematically equivalent provided the transformations 
are at least locally invertible (proving a global invertibility is a 
hard task) and nonsingular. Equivalence means that the space of 
solutions in one frame 
is in a one--to--one correspondence to the space of solutions in 
another frame. The corresponding solutions are different and physical 
quantities made up of them are different, most notably energy is very 
sensitive to various transformations. Thus dynamical equivalence of 
frames implies their physical inequivalence. As long as the theory is 
closed, i.e.~all that exists is contained in a Lagrangian of the theory 
(in the case discussed here "everything that exists" is pure gravity and 
there is no matter), this inequivalence is irrelevant as undetectable. 
All the frames are equally physical. For example, spacetime intervals 
between a given pair of events are different in distinct frames and the 
differences cannot be measured without external rods and clocks. And to 
measure energy of the gravitational triplet one needs an external device 
which is not included in the Lagrangian, yet in a closed theory neither 
an external observer nor external device does exist. In the same way 
spacetime intervals between a given pair of events are different in 
different frames and their differences cannot be measured without 
external rods and clocks. In this sense all 
the theories of physics are open: the observer and his equipment is not 
described by a tested theory\footnote{This is not so trivial as it may 
seem. There are some tendencies in quantum gravity to regard it as a 
closed theory.}. To make an NLG theory open it is necessary to couple 
it to ordinary matter and predict then some effects which may be 
observed by an external agent. \\

Coupling of matter to gravity should proceed in the same way as in 
general relativity where, however, no ambiguity appears since there is 
only one frame. For concreteness we consider now the restricted NLG 
theories, $L=f(R)$, since we will be dealing with them in Paper II. 
Clearly the coupling of matter to gravity is the 
same in any NLG theory. One takes pure gravity and chooses a frame 
consisting of a tensor $\gamma_{\mu\nu}$ regarded as a spacetime metric 
and a scalar $\phi$; these quantities are some functions of the variables 
$g_{\mu\nu}$ and $\chi$ forming HJF. The corresponding Lagrangian may 
have almost arbitrary form $L_g(R(\gamma), \phi)$ \cite{ABJT, MS1}. 
For a given kind of matter $\Psi$ its Lagrangian is constructed in 
special relativity and gets some form $L_m(\eta_{\mu\nu}, \Psi, 
\partial \Psi)$. By definition, $\gamma_{\mu\nu}$ is the metric of 
the physical spacetime and any matter \emph{minimally\/} couples to it, 
hence the matter Lagrangian for $\Psi$ becomes $L_m(\gamma_{\mu\nu}, \Psi, 
\nabla \Psi)$, where $\nabla$ is the Levi--Civita connection for 
$\gamma_{\mu\nu}$. The scalar gravity $\phi$ does not couple to the 
matter and the total Lagrangian is just $L_t = L_g +L_m$. The chosen 
frame is \emph{the physical frame\/} due to the minimal coupling. In 
all other frames where the transformed tensor playing the role of a metric 
is different from $\gamma_{\mu\nu}$, the matter is nonminimally coupled 
to it and a coupling to $\phi$ may appear. Dynamical equivalence of 
various frames remains preserved in presence of any matter while these 
frames should be regarded as unphysical since experimental devices 
measure  quantities made up of variables of the physical frame. For 
example, optical observations disclose that the light of distant 
galaxies is redshifted, what is interpreted as that expanding 
Robertson--Walker spacetime forms the physical frame, while one may 
make all computations in a conformally related frame where the spacetime 
is flat. In this case outcomes of the computations must be transformed 
back to the physical frame, the R--W spacetime, if they are to be 
confronted with observations. There is nothing new in this, the same is 
always done in classical mechanics: first one determines (experimentally) 
physical positions and momenta of a given system, makes a canonical 
transformation mixing these quantities to new variables in which the 
Hamilton equations are easiest solvable and finally makes the inverse 
transformation to express a given solution in physical variables. In 
classical mechanics this is obvious, in a gravity theory it is not. \\

Assume, as most authors applying NLG theories to cosmology actually do, 
that JF is physical, then the total Lagrangian is $L_t = f(R) + 
L_m(g_{\mu\nu}, \Psi, \nabla \Psi)$ with $\nabla$ being now the metric 
connection for $g_{\mu\nu}$. The Helmholtz Lagrangian in HJF reads 
\begin{equation}
L_H = p[R(g) - r(p)] + f(r(p)) + L_m(g_{\mu\nu}, \Psi, \nabla \Psi),
\end{equation}
from now on the spin--0 gravity in HJF is denoted $p$ and defined as 
$p\equiv \frac{\ud f}{\ud R}$ in conformity with \cite{MFF1, MS1}. 
The Lagrange equations are then $R(g) = r(p)$, 
\begin{equation}
G_{\mu\nu}(g) = \theta_{\mu\nu}(g,p) + \frac{1}{p} t_{\mu\nu}(g, \Psi)
\end{equation}
\begin{equation}
\mathrm{and} \qquad \frac{\delta L_m}{\delta \Psi}=0.
\end{equation}
Here $r(p)$ is a (possibly unique) solution of the equation 
$\frac{\ud f(r)}{\ud r}=p$ and $\theta_{\mu\nu}$ is an effective 
energy--momentum tensor for $p$ \cite{MS1} while $t_{\mu\nu}(g, \Psi)$
is the standard variational energy--momentum tensor for matter derived 
from $L_m$ with the aid of $g_{\mu\nu}$. A propagation equation for the 
scalar is derived, as previously for $\chi$, from the tensor 
$\theta_{\mu\nu}$ by taking the trace of (5) and it reads 
\begin{equation}
\Box p -\frac{2}{3}f(r(p)) +\frac{1}{3}p r(p) = 
-\frac{1}{3}g^{\mu\nu} t_{\mu\nu}(g, \Psi).
\end{equation}
It is easy to see that the scalar does not have its own Lagrangian and 
eq. (7) must be derived in this roundabout way. The metric field has 
two sources, the matter and spin--0 gravity. The equation of motion for 
any matter, (6), is independent of $p$, yet the matter forms a source 
term $g^{\mu\nu} t_{\mu\nu}$ for the scalar gravity. In this sense the 
field $p$ does not directly affect motions of matter and its effects 
are confined to affecting the metric via eq. (5). \\ 

The initial and/or boundary conditions for both $g_{\mu\nu}$ and $p$ 
are related to or even determined by a matter distribution. For example, 
in the solar system the matter distribution is dominated by the sun and 
both the fields are static spherically symmetric. Inside the sun the 
strength of the material source for $g_{\mu\nu}$ and $p$ is comparable 
and both the fields are there relatively strong and regular at the 
centre. Outside the sun the fields are fading towards spatial infinity. 
Notice that the scalar generates spherically symmetric corrections 
to Schwarzschild solution. Search for these corrections has resulted in 
the fact that the Lagrangian 
$R +1/R$ is ruled out by measurements of the PPN parameter $\gamma$ 
in the solar system \cite{ESK} and provides very stringent bounds on 
the size of possible corrections to the Einstein--Hilbert Lagrangian 
$L=R$ \cite{CB1, CST, APT}. This is an indication that Jordan frame is 
unlikely to be the physical frame. \\

If instead, Einstein frame is regarded as physical, the Helmholtz 
Lagrangian takes a form well known from general relativity, 
\begin{equation}
\tilde{L}_H = \tilde{R}(\tilde{g}) -\tilde{g}^{\mu\nu}\phi_{,\mu} 
\phi_{,\nu} - 2 V(\phi) +L_m(\tilde{g}, \Psi, \tilde{\nabla}\Psi), 
\end{equation}
where $\phi \equiv \sqrt{\frac{3}{2}} \ln p$, $V$ is a potential 
determined by $f$ and matter is minimally coupled to $\tilde{g}_
{\mu\nu}$. The Lagrangians $L_m$ in (4) and (8) have the same 
dependence on the spacetime metric, $g_{\mu\nu}$ and 
$\tilde{g}_{\mu\nu}$, respectively. As a consequence the matter 
energy--momentum tensor in EF is just $t_{\mu\nu}(\tilde{g}, 
\Psi)$. The field equations are now directly derived as the 
variational ones, 
\begin{equation}
\tilde{G}_{\mu\nu}(\tilde{g}) = T_{\mu\nu}(\tilde{g} ,\phi) + 
t_{\mu\nu}(\tilde{g}, \Psi)
\end{equation}
\begin{equation}
\stackrel{\sim}{\Box}\!\!\phi = \frac{\ud V}{\ud \phi}
\end{equation}
\begin{equation}
\mathrm{and} \qquad \frac{\delta L_m}{\delta \Psi}=0.
\end{equation}
Obviously one obtains eq. (11) from (6) by replacing $g_{\mu\nu}$ with 
$\tilde{g}_{\mu\nu}$. Now the scalar gravity is completely decoupled 
from matter and solely interacts with the metric field. This implies 
that initial and boundary conditions for $\phi$ are independent of 
matter distribution. The metric has two independent sources and its 
symmetries, boundary and initial conditions are determined by both or 
by the source that dominates. For example, in the solar system the 
$\phi$ field need not be spherical. From the spherical symmetry of the 
local spacetime one infers that the matter (the sun) is dominant on the 
RHS of eq. (9) and any small deviations from Schwarzschild metric should 
be nonspherical. The scalar field has no local (matter or any other) 
sources and fills the entire universe and approximately is described by 
one global solution. The solution is approximate since the scalar 
interacts and is affected by the spacetime metric and the latter is 
affected by local matter inhomogeneities. This global solution which is 
realized in our universe was determined by some initial and boundary 
conditions near the Big Bang. From the observations showing that the 
cosmic space is homogeneous and isotropic on large scales it is 
inferred that either the $\phi$ field is homogeneous (and time 
dependent) throughout the spacetime since the Big Bang or it is 
inhomogeneous and undetectably (even using cosmological data) weak. 
Clearly the first possibility is more attractive as it may a priori 
account for the dark energy. In any case a solution for the scalar has 
been chosen once for the universe and unlike the metric field it cannot 
be locally varied according to local (arbitrary) conditions. In other 
words, even in vicinity of a black hole, where the spacetime is extremely 
distinct from the Robertson--Walker one, the $\phi$ field only to some 
extent deviates from its overall cosmological solution. We emphasize 
that if some nonspherical deviations from the Schwarzschild metric are 
detected in the solar system (after subtracting all effects of planets, 
the Kuiper belt, the Oort cloud etc.), they may be accounted for by the 
scalar component of gravity in Einstein frame. \\

Both the frames are experimentally distinct. Which of them (if any) is 
physical? In other terms, which metric is minimally coupled to matter? 
The problem arises in any theory in which various frames appear. For 
example, in string theory in the low energy field--theory limit of string 
action one may use either the string (Jordan) frame in which the stringy 
matter is minimally coupled to the metric while the dilaton field is 
nonminimally coupled to it or transform to the conformally related Einstein 
frame where the dilaton is minimally coupled to the new metric and has 
the canonical kinetic term. The two frames are usually considered as 
completely equivalent for describing the physics of the massless modes 
of the string. This is particularly noticeable in the pre--Big Bang 
inflationary string cosmology. A superinflationary solution in string 
frame becomes an accelerated contraction in Einstein frame and vice versa. 
This drastic difference in behaviour of the cosmic scale factor in both 
the frames is irrelevant for string cosmology. In fact, the number of strings 
per unit of string volume is decreasing in time during the pre--Big Bang 
inflation in both frames and the temperature of the string gas grows in 
comparison to the temperature of the photon gas in both frames \cite{GV}. 
Whether $a(t)$ is expanding or contracting, the horizon/flatness problem 
of cosmology may be solved in each frame. \\
This frame independence for the physical effects of inflationary solutions 
in string cosmology, showing invariance of physics under local field 
redefinitions, has been found, however, only for a limited number of 
observables among those which can be constructed in this theory. Other 
observables are frame dependent. Even in the early universe spacetime 
intervals are measurable, at least in principle if not in practice, and 
these quantities depend on whether $a(t)$ grows or decreases. The very fact 
that we observe that both the optical spectra of distant galaxies and the 
cosmic microwave background radiation are redshifted, indicates that 
cosmology is not frame independent. The pre--Big Bang era of string 
cosmology is so far from us and exotic that very few effects may be 
observed today and these turn out to be frame invariant. Yet it is 
doubtful if string theory as such is frame (i.e.~conformally) 
invariant\footnote{In string theory, which is still far from making 
concrete physical predictions, it is difficult to establish what is 
measurable. Some authors claiming complete frame invariance of the theory 
seem to confuse the dynamical equivalence of various frames, which is 
indisputable, with physical equivalence.}. This would mean that the theory 
is closed. \\

In metric NLG theories which are closer to experimental physics and 
observational astronomy, the fact that most observables are frame 
dependent\footnote{Nevertheless in Jordan and Einstein frames many 
physical quantities are the same, e.g.~for black holes all the 
thermodynamical variables do not alter under a suitable Legendre map 
\cite{KM}.}, i.e.~only one frame is physical (is unique up to trivial 
field redefinitions), is hardly arguable (see however references in 
\cite{MS1}). The ultimate decision of which frame is physical will 
be given by experiment, but it should not be expected very soon. 
Before making any application of these theories it is 
necessary to establish by some theoretical arguments or by mere 
assumption which frame is measurable. More than ten years ago we studied 
in detail the full network of relationships between restricted NLG theories, 
scalar--tensor gravity theories and general relativity and effects of 
introducing matter in various frames and on this basis we gave arguments 
in favour of Einstein frame \cite{MS1}. After our work there was much 
discussion in the literature on the subject \cite{CRM}. Since the 
discovery that a modified gravity might replace the dark energy most 
authors have preferred the Jordan frame as physical. As it is 
motivated in a paper, "if one wants to consider modifications of 
gravity like scalar--tensor theory or metric $f(R)$ gravity, the Jordan 
frame should be assumed to be the physical one" since if Einstein frame 
is physical "the resulting theory will be no different from general 
relativity". These authors seem to be unaware that also in 
Helmholtz--Jordan frame the field equations for the metric tensor are 
Einstein ones. The assumption that by choosing Jordan frame as physical 
(and minimally coupling all matter in this frame) one gets a theory which 
is essentially more different from general relativity than in the other 
case, is a mere illusion. \\

We do not wish to enter the debate again. In our opinion all relevant 
arguments in favour of JF and EF have already been expounded. We only 
point out that in most cases the Einstein frame is computationally 
advantageous (and this is explicitly or implicitly acknowledged by most 
authors who implicitly or explicitly assume Jordan frame as physical). 
For example, the Zeroth Law and the Second Law of black hole 
thermodynamics for a polynomial $f(R)$ have been proved only in Einstein 
frame \cite{JKM}. On the other hand we keep open mind for the 
possibility that the real world might be not so simple as we expect. It 
might be so that the real physics is in Jordan frame while Einstein 
frame remains advantageous both in solving equations of motion and, 
what is most important, in proving general features of a given theory. 
The issue will be resolved by experiment. \\

The reader may find these comments on physical interpretation of various 
frames as unnecessary. We explain that we make these comments explicitly 
and in some length since for many researchers in cosmology these things 
are far from being clear. The level of confusion may be best seen from the 
introduction to the work \cite{CNOT} where the authors express their 
surprise that the following procedure: i) transform the equations of 
motion from Jordan frame (which is assumed to be physical) to Einstein 
frame, ii) solve the transformed equations and choose these solutions 
which seem to be physically significant and iii) transform back the 
chosen solutions to JF, does not provide physically meaningful results. 
They even quote a paper where it is claimed that "passing from one frame 
to the other can change the stability of the solution" --- a statement 
which for a mathematically oriented reader clearly means that the 
transformation is singular. The authors of \cite{CNOT} summarize their 
doubts in the question "is it correct to obtain a solution in a frame 
and then interpret it in another frame?". To leave no space for any 
ambiguity we respond: yes it is, in the same sense as it is done in 
classical mechanics where we deal with canonical transformations.\\ 

We emphasize that for the purposes of the present work the issue of 
which frame is physical, is irrelevant. We claim that the order of 
investigations should be as follows. First one chooses (on some basis) 
one or a class of NLG theories. Second, one verifies if the chosen theory 
is viable from the viewpoint of classical field theory. Third, a physical 
frame is assumed. Then equations of motion are solved either in JF or 
in HJF, EF or some other frame, depending on computational facilities 
(usually EF is most convenient). Finally the solutions should be 
transformed back to the physical frame (if were found in another one) 
to construct physical observables which will be confronted with 
observations. We stress that the second step cannot be passed over 
otherwise there is danger that the work will be wasted. In fact, in 
some papers long and nontrivial computations have been performed in a 
framework of a theory which is definitely untenable. Viability 
criteria are independent of the choice of the physical frame. This 
frame independence does not mean that whether a given theory is 
physically viable or not can be established in any frame. On the 
contrary, all the methods developed up to now to investigate the 
viability do work in Einstein frame and usually (besides one case 
mentioned in Paper II) do not work in most other 
frames. In this sense Einstein frame is mathematically distinguished. 
Yet these methods neither prove that this frame is physical nor assume 
it. We shall use Einstein frame for checking viability various $L=
f(R)$ theories. Therefore the first, basic assumption or criterion a 
theory should pass is the existence of Einstein frame: the Legendre 
transformation from JF to EF must be regular (in a neighbourhood of a 
candidate ground state solution). Once a theory meets the criteria and 
is regarded tenable one may choose the physical frame. We shall see that 
there are infinitely many viable theories and untenable ones are equally 
numerous. Obviously of all viable restricted NLG theories only one (if 
any) is correct, i.e.~will be fully confirmed by experiment and 
observations. By the time this occurs one may view any viable theory as 
a candidate for describing gravitational interactions in the nature. \\

Out of all possible viability criteria for a classical field theory the 
most appropriate (and effectively applicable) one in the case of metric 
NLG theories is the existence 
of a stable maximally symmetric ground state. Paper II is devoted to 
studying and applying this criterion. 

\section{Conclusions}
It is reasonable to conjecture that the accelerated expansion of the universe 
is not driven by an extremely exotic and unknown to physics kind of matter with 
negative pressure but is rather due to some modification of gravitational 
interactions. However in search for a modified gravity theory great caution is 
necessary. Gravitational physics is exceptional among all branches of physics 
in that there is a great variety  of competing theories, all of which are some 
variations of Einstein's general relativity. Modifications may go in all 
possible directions while in most cases gravity theories applied to cosmology 
differ from general relativity only in one axiom: the form of the field 
equations. The assumption that a gravitational Lagrangian is an arbitrary 
function $L=f(g_{\mu\nu}, R_{\alpha\beta\mu\nu})$ gives rise to 
infinity of theories and choosing the correct one is a hard task. It is 
almost invariably attempted to make this choice employing the cosmic scale 
factor $a(t)$ in the spatially flat Robertson--Walker spacetime. This 
approach is actually hopeless even in the framework of restricted metric 
gravity theories with $L=f(R)$. It has been shown \cite{CCT} that if one 
knows the exact analytic form of $a(t)$ and a number of initial conditions, 
then the function $f(R)$ is a unique solution of a linear third order ODE. 
This method does not work in practice since the equation is intractably 
complicated. The fundamental cause that the method cannot work is that the 
real physical spacetime is not R--W one. The cosmic matter distribution 
becomes homogeneous and isotropic only asymptotically at large scales. If 
one were ingenious enough to solve the differential equation for $f(R)$ 
employing that form of $a(t)$ which best fits all the astronomical data, 
the resulting Lagrangian would be rather different from the true one. The 
standard Friedmannian cosmology based on general relativity is the best fit 
to the large scale properties of the universe (besides its acceleration) 
but not vice versa. The standard model (and any other model too) provides 
only some approximation to the reality and a slightly modified 
approximation, which also fits the observations with a satisfactory 
accuracy, would lead to a gravity theory different from general relativity. 
It has recently been found that an attempt to reconstruct the 
$\Lambda$CDM cosmological model as a $L=f(R)$ gravity provides the required 
$L=R-2\Lambda$ as one of a number of solutions but not as the one most 
probable (i.e.~it is not the solution corresponding to attractor 
points of an autonomous system) \cite{FNP}. In short: from an 
approximate solution it is impossible to reconstruct in a reliable way 
the correct equation of motion.\\

On the other hand if the exact form of the spacetime metric accurately 
corresponding to the distribution of the cosmic matter were known, the 
problem of how to reconstruct the underlying Lagrangian would be open: the 
method developed in \cite{CCT} is specific to R--W spacetime and does not 
work in other cases. We stress that, contrary to a common belief, the R--W 
spacetime is particularly deceptive and unsuitable for recontructing the 
underlying Lagrangian. This spacetime has a high symmetry and is `flexible' 
in the sense that it contains an arbitrary function, so that it is a solution 
in \emph{any\/} metric gravity theory (while Minkowski space is not). Hence 
for fundamental rather than technical reasons it should not be used for the 
reconstruction. If one believes at all that it is possible to recover the 
Lagrangian from one (empirically found) solution, one should apply a 
solution which does not appear as such in most of gravity theories and is 
a characteristic feature of a possibly narrow class of theories. \\

A direct comparison of predictions of a given theory with observations is 
obstructed by the fact that any nonlinear gravity theory may be formulated 
in infinite number of distinct frames and many of them have advantage over 
the original Jordan frame (in which all the theories are initially 
formulated) in displaying the number of degrees of freedom, the particle 
spectrum and the dynamics of these fields. Since the problem of which 
frame is physical (i.e.~consists of directly measurable dynamical 
variables) still remains a matter of a vivid debate and since for this 
reason any agreement (or disagreement) of the given theory with the 
observational data may be criticized, instead of attempting to deduce a 
gravity theory from the data and prior to attempting such a confrontation 
for a chosen theory, one should verify if the theory meets the general 
requirements imposed on a classical field theory. A general 
$L=f(g_{\mu\nu}, R_{\alpha\beta\mu\nu})$ metric theory has eight degrees 
of freedom and describes a gravitational triplet consisting of the metric, 
a massive spin--two field and a massive scalar field. For $L=f(R)$ the 
spin--2 field disappears. All the $L=f(g_{\mu\nu}, R_{\alpha\beta})$ 
theories less differ from each 
other than it is expected in the original Jordan frame since they may be 
mapped onto general relativity (including the two massive fields) and in 
this sense the latter theory is clearly distinguished as being a universal 
Hamiltonian image of all these theories.\\ 
How many conditions a gravity theory should 
satisfy to be regarded a viable one may be disputable (e.g.~ should it 
have quantization properties better than general relativity?). The 
criterion that a theory have a stable ground state being a maximally 
symmetric spacetime is indisputable. This criterion is investigated in 
the subsequent Paper II. It is shown there that there is 
an infinity of viable gravity theories. One should therefore apply 
other viability criteria coming from classical field theory. The most 
natural one would be the existence of a proper Newtonian limit. 
However at present it is difficult to formulate this condition in a 
precise and effective way. And we stress that at the present level of 
knowledge it is impossible to find out a system of selection rules 
allowing one to choose a unique theory out of the whole set of the 
metric gravity theories. Therefore above all one should provide a 
deeper physical motivation (different from mere applying a number of 
selection rules and from the wish to account for the 
cosmic acceleration) for choosing a specific Lagrangian rather than any 
other. In other terms the cosmic acceleration should be a new important 
test for a modified gravity theory but it does not provide a way for 
reconstructing it.   \\

\section*{Acknowledgments}
 
I am grateful to Michael Anderson, Piotr Bizo\'n, Piotr Chru\'sciel, 
Helmut Friedrich, Zdzis\l{}aw Golda and Barton Zwiebach for extensive 
discussions, helpful comments and explanations.
This work is supported in part by a Jagellonian University grant.


\begin{thebibliography}{}
\frenchspacing

\bibitem{Stel1}
K. S. Stelle,
Phys. Rev.  \textbf{D16} (1977) 953. 

\bibitem{Starob}
A. A. Starobinsky, Phys. Lett. \textbf{91B} (1980) 99;
R. Kerner, Gen. Rel. Grav. \textbf{14} (1982) 453;
A. A. Starobinsky and H.-J. Schmidt, Class. Quantum Grav. \textbf{4} 
(1987) 695.

\bibitem{Carr1}
S. M. Carroll, V. Duvvuri, M. Trodden and M. S. Turner, Phys. Rev. 
\textbf{D70} (2004) 043528. 

\bibitem{Carr2}
S. M. Carroll, A. De Felice, V. Duvvuri, D. A. Easson, M. Trodden 
and M. S. Turner, Phys. Rev. \textbf{D71} (2005) 063513.

\bibitem{NO1}
S. Nojiri and S. D. Odintsov, Phys. Rev. \textbf{D68} (2003) 123512. 

\bibitem{NO2}
S. Nojiri and S. D. Odintsov, Phys. Rev. \textbf{D74} (2006) 086005 
[hep-th/0608008].

\bibitem{CST}
S. Capozziello, A. Stabile and A. Troisi, Mod. Phys. Lett. \textbf{A21} 
(2006) 2291 [gr-qc/0603071].

\bibitem{Cem}
J. A. R. Cembranos, Phys. Rev. \textbf{D73} (2006) 064029. 

\bibitem{BBH}
A. W. Brookfield, C. van de Bruck and L. M. H. Hall, 
Phys. Rev. \textbf{D74} (2006) 064028, [hep-th/0608015].

\bibitem{Ren}
A. D. Rendall, Class. Quantum Grav. 
\textbf{24} (2007) 667, [gr-qc/0611088]; Class. Quantum Grav. 
\textbf{21} (2004) 2445.

\bibitem{MV}
T. Multam\"aki and I. Vilja, Phys. Rev. 
\textbf{D74} (2006) 064022, [astro-ph/0606373].

\bibitem{ESK}
A. L. Erickcek, T. L. Smith and M. Kamionkowski, 
Phys. Rev. \textbf{D74} (2006) 121501, [astro-ph/0610483].

\bibitem{KKL}
V. Kagramanova, J. Kunz and C. L\"ammerzahl, Phys. Lett. 
\textbf{B634} (2006) 465. 

\bibitem{Olm}
G. J. Olmo, Phys. Rev. Lett. \textbf{95} (2005) 261102; Phys. Rev.  
\textbf{D72} (2005) 083505; Phys. Rev. \textbf{D75} (2007) 023511, 
[gr-qc/0612047]. 

\bibitem{RI}
M. L. Ruggiero and L. Iorio, JCAP \textbf{0701} (2007) 010, 
[gr-qc/0607093].

\bibitem{APT}
L. Amendola, D. Polarski and S. Tsujikawa, 
Phys. Rev. Lett. \textbf{98}: 131302, 2007. 


\bibitem{Ch1}
T. Chiba, Phys. Lett. \textbf{B575} (2003) 1. 

\bibitem{CSE}
T. Chiba, T. L. Smith and A. L. Erickcek, [astro-ph/0611867]. 

\bibitem{CB1}
T. Clifton and J. D. Barrow, 
Phys. Rev.  \textbf{D72} (2005) 103005. 

\bibitem{NVA}
I. Navarro and K. van Acoleyen, 
JCAP \textbf{0702}: 022, 2007  [gr-qc/0611127]. 

\bibitem{DBD}
S. Das, N. Banerjee and N. Dadhich, Class. Quantum Grav. 
\textbf{23} (2006) 4159.

\bibitem{CDD}
A. de la Cruz-Dombriz and A. Dobado, Phys. Rev. \textbf{D74} 
(2006) 087501, [gr-qc/0607118].

\bibitem{Ea}
D. A. Easson, 
JCAP \textbf{0702}: 004, 2007 [astro-ph/0608034].

\bibitem{Woo}
R. P. Woodard, \textit{Proceedings of 3rd Aegean Summer School 
"The Invisible Universe: Dark Matter and Dark Energy", 
September 2005}, [astro-ph/0601672]. 

\bibitem{BBPST}
R. Bean, D. Bernat, L. Pogosian, A. Silvestri and M. Trodden, 
Phys. Rev.  \textbf{D75} (2007) 064020. 

\bibitem{MSW}
O. Mena, J. Santiago  and J. Weller, 
Phys. Rev. Lett. \textbf{96} (2006) 041103.

\bibitem{AEMM}
M. Amarzguioui, O. Elgaroy, D. F. Mota and T. Multam\"aki, Astron. 
Astrophys. \textbf{454} (2006) 707, [astro-ph/0510519]. 

\bibitem{CCT}
S. Capozziello, V. F. Cardone and A. Troisi, Phys. Rev. 
\textbf{D71} (2005) 043503. 

\bibitem{FNP}
S. Fay, S. Nesseris and L. Perivolaropoulos, [gr-qc/0703006].

\bibitem{Di}
R. Dick, 
Gen. Rel. Grav. \textbf{36} (2004) 217. 

\bibitem{BB}
A. J. Bustelo and D. E. Barraco, Class. Quantum Grav. 
\textbf{24} (2007) 2333  [gr-qc/0611149].

\bibitem{BD}
D. G. Boulware and S. Deser,
Phys. Rev.  \textbf{D6} (1972) 3368. 

\bibitem{FT}
E. S. Fradkin and A. A. Tseytlin,
Nucl. Phys.  \textbf{B201} (1982) 469.

\bibitem{BCh}
N. Barth and S. M. Christensen,
Phys. Rev.  \textbf{D28} (1983) 1876. 

\bibitem{HOW1}
A. Hindawi, B. A. Ovrut and D. Waldram, 
Phys. Rev.  \textbf{D53} (1996) 5597. 

\bibitem{Sa}
R. K. Sachs, \textit{Gravitational radiation}, in:  
\textit{Relativity, Groups and Topology, Les Houches 1963}, edited by
C. DeWitt and B. DeWitt, Gordon and Breach, New York 1964, 
pp. 523--562.

\bibitem{MS1}
G. Magnano and L. M. Soko\l{}owski,
Phys. Rev. \textbf{D50} (1994) 5039, [gr-qc/9312008].

\bibitem{MFF1}
G. Magnano, M. Ferraris and M. Francaviglia, 
Gen. Rel. Grav. \textbf{19} (1987) 465. 

\bibitem{JK}
A. Jakubiec and J. Kijowski, 
Phys. Rev.  \textbf{D37} (1988) 1406; Gen. Rel. Grav. \textbf{19} 
(1987) 719. 

\bibitem{MFF2}
G. Magnano, M. Ferraris and M. Francaviglia, 
Class. Quantum Grav. \textbf{7} (1990) 557. 

\bibitem{MS2}
G. Magnano and L .M. Soko\l{}owski, Ann. Phys. (N.Y.) \textbf{306} 
(2003) 1. 

\bibitem{SS}
L. M. Soko\l{}owski and A. Staruszkiewicz, 
Class. Quantum Grav. \textbf{23} (2006) 5907. 

\bibitem{Stel2}
K. S. Stelle,
Gen. Rel. Grav. \textbf{9} (1978) 353. 

\bibitem{ABJT}
J.C. Alonso, F. Barbero, J. Julve and A. Tiemblo, Class. Quantum 
Grav. \textbf{11} (1994) 865. 

\bibitem{HOW2}
A. Hindawi, B. A. Ovrut and D. Waldram, 
Phys. Rev.  \textbf{D53} (1996) 5583. 

\bibitem{AD}
C. Aragone and S. Deser, 
Nuovo Cim. \textbf{3A} (1971) 709;
Nuovo Cim. \textbf{57B} (1980) 33.

\bibitem{Tom}
E. T. Tomboulis, 
Phys. Lett. \textbf{B389} (1996) 225.

\bibitem{Wa}
R. M. Wald, 
Class. Quantum Grav. \textbf{4} (1987) 1279. 

\bibitem{Ch2}
T. Chiba, 
JCAP \textbf{0503} (2005) 008. 

\bibitem{GV}
M. Gasperini and G. Veneziano, 
Phys. Reports  \textbf{373} (2003) 1;
E. Alvarez and J. Conde, Mod. Phys. Lett. \textbf{A17} (2002) 413.

\bibitem{KM}
J. Koga and K. Maeda, 
Phys. Rev.  \textbf{D58} (1998) 064020. 

\bibitem{CRM}
S. Capozziello, E. de Ritis and A. A. Marino, 
Class. Quantum Grav. \textbf{14} (1997) 3243;
R. Dick, Gen. Rel. Grav. \textbf{30} (1998) 435;
R. Casadio and B. Harms, Mod. Phys. Lett. \textbf{A14} (1999) 1089;
V. Faraoni and E. Gunzig, Int. J. Theor. Phys. \textbf{38} (1999) 
217;
V. Faraoni, E. Gunzig and P. Nardone, Fund. Cosmic Phys. \textbf{20} 
(1999) 121;
A. Mac\'ias and A. Garc\'ia, Gen. Rel. Grav. \textbf{33} (2001) 889;
D. N. Vollick, Class. Quantum Grav. \textbf{21} (2004) 3813;
E. E. Flanagan, Class. Quantum Grav. \textbf{21} (2004) 3817.

\bibitem{JKM}
T. Jacobson, G. Kang and R. C. Myers, 
Phys. Rev.  \textbf{D52} (1995) 3518. 

\bibitem{CNOT}
S. Capozziello, S. Nojiri, S. D. Odintsov and A. Troisi,  
Phys. Lett. \textbf{B639} (2006) 135.

\end{thebibliography}
\end{document}